\documentstyle[preprint,aps]{revtex}
\tightenlines
\begin{document}
\draft

\title{Classical and Quantum Coherent State Description of $N\bar{N}$
Annihilation at rest in the Skyrme Model with $\omega$ Mesons}

\author{Bin Shao and R. D. Amado}
\address{Department of Physics,
University of Pennsylvania, Philadelphia,
PA 19104, USA}

\date{\today}

\maketitle

\begin{abstract}
We model the $N\bar{N}$ system at rest as a baryon number zero lump
in the Skyrme model including the $\omega$ field. We integrate the classical
equations of motion from the highly non-perturbative annihilation region
into the non-interacting radiation zone. Subsequent coherent state
quantization of the radiation field gives a good description of
the pion spectrum from annihilation at rest.
\end{abstract}

\newpage
\section{Introduction}

We have recently shown that
low energy nucleon-antinucleon annihilation
directly into pions
can be well described as occurring
through a coherent classical pion wave that
emerges very quickly from the annihilation region
\cite{Amado1}, \cite{Amado2}.
The physical, quantum pions detected from
annihilation can be described by
quantizing this classical pion
burst using the method of coherent states
\cite{Glauber}. We project
these coherent  states
onto states of fixed four-momentum \cite{HornSilver}
and definite isospin \cite{Botke}. A correct picture of
the pion spectrum, pion number probability and charge
averages emerges from this simple picture. Our approach
is inspired  by calculations of Skyrmion anti-Skyrmion
annihilation, where it is found that the classical
Skyrmion fields annihilate into a burst of pion radiation
nearly as soon as the Skyrmion and anti-Skyrmion touch,
and that that pion wave takes away the baryon number and energy
as fast as causality will permit \cite{Sommermann}, \cite{SWA}.
This picture, and our first treatment
dealt only with the direct pions and did not
address the roughly $40\%$ of annihilations that go first
through meson resonances, mostly omega and rho mesons, and
then to pions \cite{Amsler}. The purpose of this paper is to
begin to address this deficiency  by including omega mesons
in our description of annihilation.  We
will return to rho mesons in a subsequent paper.

In our previous work \cite{Amado1},\cite{Amado2}, we approached
the classical pion wave from annihilation phenomenologically,
and concentrated on projections of the coherent state.  We found
that a very simple parameterization of the wave source gave
a good description of the pion spectrum. The  only parameter
we needed was the overall amplitude of the wave, and this we
could fit to the total energy released.  When coupling to omega
mesons the situation is more complicated, since both the relative
amplitude and shape of the omega and pion waves, even classically,
should be dynamically determined. Thus we need to make a dynamical
theory of the coupled  pion and omega
fields. Again we do that classically. We construct a classical picture
of the annihilation proceeding into interacting pion and
omega fields. We propagate  those fields, classically, into the
radiation zone, where they no longer interact,
and then quantize the radiation into coherent states
for the pions and omegas separately.  Finally we allow the
omega mesons to decay into pions, assuming that that radiation is incoherent
with respect to the direct pions.  This approach of treating the
non-perturbative aspects of QCD classically and then quantizing
the subsequent radiation in the radiation zone using the methods of
coherent states has recently appeared in a number of applications
\cite{Gong},\cite{Gould}.

For a dynamical theory of coupled, classical pion and omega fields,
we use the Skyrme model modified to include omegas \cite{adkins}.
Recall that the usual Skyrme lagrangian \cite{Skyrme}
describes a classical pion field only, and that baryons emerge
as topologically stable configurations of that pion field.  Also
recall that this theory is a candidate for what QCD looks like in
the classical or large number of colors limit. To include
omega mesons in
this formalism, the classical
omega field is coupled to the baryon current. This
helps to stabilize the Skyrmion.  To model annihilation
at rest in this formalism, we follow
our earlier work \cite{SWA} and begin with a
spherically symmetric ``blob" of Skyrmionic matter of size
about 1 fm, baryon number zero and the total energy of
two nucleons.  This initial configuration is not a
static solution of the Skyrme equations, but rather evolves
according to those equations into
radiating and interacting pion and omega
fields. As these fields radiate outwards,
they diminish like $1/r$,
decouple and become radiating free fields.
We use these radiation fields to  construct coherent states
for the omegas and pions. We then
project these coherent states
onto states of good four-momentum and isospin,
as before, and examine
the consequences for nucleon-antinucleon phenomenology.
We find very good agreement with the principal features of
annihilation, now including the omega channel.
The coherent
state picture seems to work in spite of the fact that the mean
number of omegas is of order one.
We assume that omega decay occurs outside the interaction
region of the classical fields and therefore
we do not treat the pions from omega decay as
coherent with the direct pions.
But we do, of course, treat the pions from the omega as
part of the observed pion spectrum.

In Section II we present  the classical Skyrme lagrangian with
coupling to the omega field and the dynamical equations for
propagation of the fields from a spherically symmetric
source.  We also show the classical field configurations
that emerge from the decay of an initial ``blob" meant to
model annihilation.  Section III discusses the quantization
of the classical $\pi$ and $\omega$ fields using the method
of coherent states modified to include projection onto
fixed isospin and four-mometum.  Section IV gives the  results
of this formalism for the pion spectrum from annihilation.
We find that the mean number of pions, the variance, and the
entire shape of the pion number spectrum closely resemble experiment
\cite{Dover}.
In our previous, pions only, calculation \cite{Amado2},
we found that only an even number of pions could come from
even isospin states and only an odd number from odd.
The pions from omega decay remove
this artifact, giving smooth pion number probabilities
in each isospin channel.  Section V presents some conclusions
and directions for future work.

\vspace*{0.5cm}
\section{Classical Calculation}

We begin with a classical theory of the coupled pion and omega
fields based on the Skyrme model. This coupled theory was written
down by Adkins and Nappi \cite{adkins}. In terms of the usual SU(2)
valued unitary field $U$ of the Skymre picture
the lagrangian is given by
\begin{equation}
{\cal L}={\cal L}_{\pi}+{\cal L}_{\omega}+{\cal L}_{int}
\end{equation}

\begin{equation}
{\cal L}_{\pi}=-\frac{f_{\pi}^{2}}{2}Tr\left( \partial_{\mu}U
\partial^{\mu}U^{\dagger}\right)+\frac{1}{2}m_{\pi}^{2}f_{\pi}^{2}Tr(U-1)
\end{equation}

where
\begin{equation}
U=\exp( \frac{i}{f_{\pi}} \vec{\tau}\cdot\vec{\pi} )
\end{equation}

\begin{equation}
{\cal L}_{\omega}=-\frac{1}{4}\omega_{\mu\nu}\omega^{\mu\nu}+\frac{m^{2}}{2}
\omega_{\mu}\omega^{\mu}
\end{equation}

\begin{equation}
{\cal L}_{int}=\beta\omega_{\mu}B^{\mu}
\end{equation}

\begin{equation}
B^{\mu}=\frac{1}{24\pi^{2}}\epsilon^{\mu\nu\alpha\beta}
Tr\left[ (U^{\dagger}\partial_{\nu}U)(U^{\dagger}\partial_{\alpha}U)
(U^{\dagger}\partial_{\beta}U)\right]
\end{equation}
The first term, ${\cal L}_{\pi}$, is the standard non-linear sigma model
term with a pion mass term included.  It has two paramenters, $f_{\pi}$,
the pion decay constant, which it is customary to vary a little to
fit the nucleon mass, and the pion mass, which we do not vary. Note that
there is no Skyrme term in the lagrangian, since the coupling to the
omega now stabilizes the theory. The second term, ${\cal L}_{\omega}$,
is the free omega lagrangian, where $\omega_{\mu}$ is the omega field,
$\omega_{\mu \nu}$ the corresponding antisymmetric field strength tensor,
and $m$ the omega mass. The interaction term, ${\cal L}_{int}$, couples
the omega field to the baryon current, $B^{\mu}$, and $\beta$ is a coupling
parameter. We take $\beta=15.6$ and $f_{\pi}=62.0\mbox{MeV}$
from the work of Adkins and Nappi \cite{adkins}.
${\cal L}_{int}$ describes the coupling of the omega to three pions.

To model annihilation, we will take for our initial configuration
a spherically symmetric ``blob" of matter  with zero baryon number
but total energy of two nucleons. Thus we only need solutions of
the lagrangian that have spherical symmetry. This is a great simplification.
For the spherically symmetric case we may write
\begin{equation}
U=\exp( i \vec{\tau}\cdot\hat{r}F(r, t) )
\end{equation}
in which case the lagrangian becomes,
\begin{eqnarray}
{\cal L}&=&\frac{f_{\pi}^{2}}{2}\left[ \left( \frac{\partial F}{\partial t}
\right)^{2}-2\frac{\sin^{2}F}{r^{2}}-\left( \frac{\partial F}{\partial r}
\right)^{2}\right]+m_{\pi}^{2}f_{\pi}^{2}(\cos F -1) \nonumber
\\ && +
\frac{1}{2}\left[ \left( \frac{\partial \omega_{r}}
{\partial t}-\frac{\partial \omega_{0}}{\partial r} \right)^{2}
+\left( \frac{\partial\omega_{\theta}}{\partial t} \right)^{2}
+\left( \frac{\partial\omega_{\phi}}{\partial t} \right)^{2}
\right]
+\frac{m^{2}}{2}\left[\omega_{0}^{2}-\omega_{r}^{2}-\omega_{\theta}^{2}
-\omega_{\phi}^{2}\right] \nonumber \\ &&
-\frac{\beta}{2\pi^{2}}\frac{\sin^{2}F}{r^{2}}
\left[\omega_{0}\frac{\partial F}{\partial r}-\omega_{r}\frac{\partial F}
{\partial t}\right]
\end{eqnarray}
As we can see, $\omega_{\theta}$ and $\omega_{\phi}$ are completely
decoupled from the $\pi$ field. We therefore shall not consider these
two components.

The energy of the system is given by
\begin{equation}
H=4\pi \int_{0}^{\infty} r^{2} dr {\cal H}(r)    \label{eq:energy}
\end{equation}
where
\begin{eqnarray}
{\cal H}(r)&=&\frac{f_{\pi}^{2}}{2}\left[
\left( \frac{\partial F}{\partial t} \right)^{2}
+\left( \frac{\partial F}{\partial r} \right)^{2}
+2\frac{\sin^{2}F}{r^{2}} \right]
+f_{\pi}^{2}m_{\pi}^{2}(1-\cos F)
+\frac{1}{2}\left[
\left( \frac{\partial \omega_{r}}{\partial t} \right)^{2}
\right.  \nonumber
\\ && \left.
-\left( \frac{\partial \omega_{0} }{\partial r} \right)^{2} \right]
+\frac{m^{2}}{2}[\omega_{r}^{2}-\omega_{0}^{2} ]
+\frac{\beta}{2\pi^{2}}\sin^{2}F
\left( \omega_{0} \frac{\partial F}{\partial r}
-\omega_{r} \frac{\partial F}{\partial t} \right).
\end{eqnarray}

The equations of motion are given by
\begin{equation}
f_{\pi}^{2} \left[\frac{\partial^{2}F}{\partial t^{2}}-
\frac{\partial^{2}F}{\partial r^{2}}-
\frac{2}{r}\frac{\partial F}{\partial r}
+\frac{\sin 2F}{r^{2}}+m^{2}_{\pi}\sin F\right]=\frac{\beta}{2\pi^{2}}
\frac{\sin^{2}F}{r^{2}}\left( \frac{\partial\omega_{0}}{\partial r}
-\frac{\partial\omega_{r}}{\partial t} \right),
\end{equation}
\begin{equation}
\left[\frac{\partial^{2}}{\partial t^{2}}-
\frac{\partial^{2}}{\partial r^{2}}-
\frac{2}{r}\frac{\partial}{\partial r}
+m^{2}\right]\omega_{0}=\frac{\beta}{2\pi^{2}}\frac{\sin^{2}F}{r^{2}}
\frac{\partial F}
{\partial r},
\end{equation}
\begin{equation}
\left[\frac{\partial^{2}}{\partial t^{2}}-
\frac{\partial^{2}}{\partial r^{2}}-
\frac{2}{r}\frac{\partial}{\partial r}
+\frac{2}{r^{2}}
+m^{2}\right]\omega_{r}
=\frac{\beta}{2\pi^{2}}\frac{\sin^{2}F}{r^{2}}\frac{\partial F}
{\partial t}.
\end{equation}

We model the initial {\em static} field configuration as
\begin{equation}
F(r,t=0)=h\frac{r}{r^{2}+a^{2}}\exp(-r/a)
\end{equation}
with $a=1/m_{\pi}$ and $h$ determined by the total energy.
This form corresponds to a compact initial ``blob" of
zero baryon number. Since we want to begin with the
system at rest we take
\begin{equation}
\dot{F}=\dot{\omega}_{r}=\dot{\omega}_{0}=0.
\end{equation}
We may also take
\begin{equation}
\omega_{r}(r,t=0)=0
\end{equation}
since $B_{r}=0$ at $t=0$.
These conditions determine $\omega_0(r,t=0)$, we find
\begin{equation}
\omega_{0}(r,t=0)=\int_{0}^{\infty}dr^{'}G(r,r^{'})\left[-\beta r^{'2}
B^{0}(r^{'})\right]
\end{equation}
where
\begin{equation}
G(r,r^{'})=\frac{1}{2mrr^{'}}\left( e^{-m|r-r^{'}|}-e^{-m(r+r^{'})} \right)
\end{equation}
Using the initial values we integrate the equations of motion
to determine the fields at later times.  As the fields propagate
outwards, they diminish in size, making the non-linear and coupling
terms in the lagrangian less important.  Thus we can define a
radiation zone where the fields propagate as linear free fields.
In the radiation zone, we can calculate the pion,
$f(k)$, and omega, $g(k)$, momentum
distribution amplitudes from the expressions
\begin{equation}
\frac{d N_{\pi}}{d^{3}k}=\left| f(k) \right|^{2}=
\frac{1}{\pi k_{0}^{\pi} }
f_{\pi}^{2} \left| \int_{0}^{\infty} dr r^{2} j_{1}(kr)
(k_{0}^{\pi} + i\frac{\partial}{\partial t}) F(r,t) \right|^{2},
\label{eq:pionDist}
\end{equation}
and
\begin{equation}
\frac{d N_{\omega}}{d^{3}k}=\left| g(k) \right|^{2}=
\frac{1}{\pi k_{0}^{\omega} }
\left| \int_{0}^{\infty} dr r^{2} j_{1}(kr)
(k_{0}^{\omega}+i\frac{\partial}{\partial t})\omega_{r}(r,t) \right|^{2},
\label{eq:omegaDist}
\end{equation}
respectively.

The results of our calculations are shown in the figures. We have fixed
$h$ in our initial configuration by the requirement that our initial
state have total energy of 2 GeV. Figure 1 shows the pion field configuration,
$F$, as a function of $r$ and $t$, both in Fermi. We see that it emerges as
a coherent pulse. Much of that pulse travels nearly along the light
cone, because of the light pion mass.  The $\omega_0$ and $\omega_r$
fields are shown in Figures 2 and 3. Recall that $\omega_r$ is initially
zero.  We see that both emerge more slowly and with more complex behavior
than the pion field, due to the large omega mass.
The following figures
show the energy density.  First we see (in Figure 4) the total
energy density then in Figures 5 and 6 the energy density in the pion field
and in the r-component of the omega field.  That is the part of the omega
that corresponds to radiation.  In all these plots we show the energy
density multiplied by $4\pi r^2$, that is the total energy at any time is
obtained from them by integration over $ dr$ only. For the total,
this integral must yield 2 GeV at every time, and we use this
test to verify the
stability of our calcuation.
Without the factor of $r^2$, the energy density would be seen to
decrease rapidly with $r$, reflecting the corresponding decrease in
the fields.  This descrease is most dramatically seen in Figure 7
where we plot the interaction energy density.  We see that past the
first few time steps, the interaction energy disappears, signifying
the decoupling of the fields.  Beyond this point not only do the
fields decouple, but, to a very good approximation, they obey
a linear wave equation.
This region is the radiation zone, and there we can use Eqs.
(\ref{eq:pionDist}) and
(\ref{eq:omegaDist}) to obtain the pion and omega momentum distributions.
In Figures 8 and 9 we plot the reduced momentum distribution densities
which are defined by
\begin{equation}
\rho_{f}(\bar{k})=4\pi m k^{2}|f(k)|^{2}
\end{equation}
and
\begin{equation}
\rho_{g}(\bar{k})=4\pi m k^{2}|g(k)|^{2}
\end{equation}
as functions of scaled dimensionless wave number $\bar{k}=k/m$.
We see that
the pion distribution is quite sharply peaked, roughly at a wave number
corresponding to the size of the initial distribution. The omega distribution
is much broader, largely because that field propagates with a mass that
is large compared with other scales in the problem.
The momentum amplitudes,
$f(k)$ and $g(k)$ are the basis for our coherent state quantization.

It is amusing to compare the dynamical calculation reported above with our
previous phenomenological description of annihilation into pions only,
\cite{Amado1}, \cite{Amado2}. We can repeat the
dynamical calculation done above using the
same initial form, $F(r,t=0)$, for the baryonic configuration,
but this time having no omega field and using the standard Skyrme
lagrangian \cite{Skyrme}. This is the appropriate dynamics for
our previous, pions only, picture.
Once again we propagate the pion field into
the radiation zone and calcuate the momentum density, $f(k)$.  The
corresponding reduced distribution densities are
shown in Figure 10, where it is compared with the phenomenological
form used in \cite{Amado1}, \cite{Amado2}.  We see that the two are
remarkably alike.  Their overall scale agrees because they were fit to the
same total energy, and they both come from an initial distribution
of about the same size, but they were arrived at very differently and
their agreement is certainly both satisfying and surprising.

\section{Coherent State Calculation}
We construct quantum coherent states for the $\omega$
and $\pi$  mesons
from the classical fields obtained in the previous section using
the same methods we used before \cite{Amado1}, \cite{Amado2}.
Recall that we wish to construct coherent states that
have fixed four-momentum and isospin. To do this we
define a
pion field operator that creates pions at the space-time position
$x$ and pointing in the isospin direction ${\hat T}$ by
\begin{equation}
F(x)=\int d^{3}k f(\vec{k}) \vec{a}^{\dagger}_{\vec{k}}\cdot\hat{T}e^{-ik\cdot
x},
\end{equation}
where in the exponent under the integral, the energy component
of $k$ is taken as the on-shell value $k_0 = \sqrt{k^2 + m_{\pi}^2}$.
The corresponding omega field operator is given by
\begin{equation}
G(x)=\int d^{3}k^{'} g(\vec{k}^{'}) b^{\dagger}_{\vec{k}^{'}}e^{-ik^{'}\cdot
x}.
\end{equation}
with appropriate definition of the fourth component of $k'$.
In these forms, $ \vec{a}^{\dagger}_{\vec{k}}$ creates an
isovector pion with
momentum ${\vec{k}}$ and  $b^{\dagger}_{\vec{k}^{'}}$ creates an isoscalar
omega with momentum ${\vec{k}^{'}}$.

To impose both energy-momentum and isospin conservations, we take
the quantum coherent state of the pion and omega
system in the radiation region to be given by
\begin{equation}
|I,I_z,K>=\int\frac{d^{4}x}{(2\pi)^{4}}\frac{d\hat{T}}{\sqrt{4\pi}}
e^{iK\cdot x}Y_{II_z}(\hat{T}) (e^{F(x)+G(x)}-1-F(x)-G(x))|0>.
\label{eq:cs}
\end{equation}
As we explain in \cite{Amado1}, we subtract the ``one'' as well as
the one particle (pion or omega) states since they are not permitted
by energy-momentum conservation and make the calculation numerically
unstable.
The states defined in (\ref{eq:cs}) are not normalized,
but they are orthogonal.
We have
\begin{equation}
<I,I_z,K|I^{'},I_{z}^{'},K^{'}>=\delta^{4}(K-K^{'})\delta_{II^{'}}
\delta_{I_z I^{'}_z}{\cal I}(K)
\end{equation}
The normalization factor is given by
\begin{equation}
{\cal I}(K)=\int \frac{d^{4}x}{(2\pi)^{4}} \frac{d\hat{T}d\hat{T}^{'}}{4\pi}
Y_{II_z}^{*}(\hat{T})Y_{II_z}(\hat{T}^{'})
e^{iK\cdot x} (e^{ \rho_{\pi}(x)\hat{T}\cdot\hat{T}^{'}
+\rho_{\omega}(x)}-1-
\rho_{\pi}(x)\hat{T}\cdot\hat{T}^{'}-\rho_{\omega}(x) )
\end{equation}
where
\begin{equation}
\rho_{\pi}(x)=\int d^{3}p |f(\vec{p})|^{2}e^{-ip\cdot x}
\end{equation}
and
\begin{equation}
\rho_{\omega}(x)=\int d^{3}p |g(\vec{p})|^{2}e^{-ip\cdot x}
\end{equation}
The normalization integral is difficult to calculate numerically, even
after the subtraction. Hence we use the expansion method
we developed before \cite{Amado2}, generalized to the
case of two meson types.  We thus get
\begin{equation}
{\cal I}(K)=\sum_{N_{\pi}+N_{\omega}\ge 2}
\frac{I(K, N_{\pi},N_{\omega})}{N_{\pi}!N_{\omega}!}F(N_{\pi},I).
\end{equation}

The probability for finding $N_{\pi}$ pions and $N_{\omega}$ omega is given by
\begin{equation}
p(N_{\pi},N_{\omega})=\frac{1}{{\cal I}(K)}
\frac{I(K, N_{\pi},N_{\omega})}{N_{\pi}!N_{\omega}!}
F(N_{\pi},I)          \label{eq:jointProb}
\end{equation}
where
\begin{equation}
I(K, N_{\pi},N_{\omega})=
\int \delta^{4}(K-\sum_{i=1}^{N_{\pi}}p_{i}-\sum_{j=1}^{N_{\omega}}
q_{j}) \prod_{i=1}^{N_{\pi}} d^{3}p_{i} |f(\vec{p}_{i})|^{2}
\prod_{j=1}^{N_{\omega}}d^{3}q_{j} |g(\vec{q}_{j})|^{2} \label{eq:phase}
\end{equation}
and
\begin{eqnarray}
F(N_{\pi},I)&=&\int \frac{d\hat{T}d\hat{T}^{'}}{4\pi}
Y^{*}_{II_z}(\hat{T})Y_{II_z}(\hat{T}^{'})(\hat{T}\cdot\hat{T}^{'})^{N_{\pi}}
\nonumber \\
&& =\left\{   \begin{array}{ll}
0 & I >  N_{\pi} \mbox{ and } I-N_{\pi} \mbox{ is odd} \\
\frac{ N_{\pi}! }
{ (N_{\pi}-I)!! (I+N_{\pi}+1)!! } & I\le N_{\pi} \mbox{ and }
I-N_{\pi} \mbox{ is even}.
\end{array}   \right.
\end{eqnarray}
If either $N_{\pi}$ or $N_{\omega}$ equal zero, the corresponding
terms will not appear in  Eq.(\ref{eq:phase}).

The mean number of $\pi$'s of isospin type $\mu$
in the state is given by
\begin{equation}
N_{\pi\mu}=\frac{1}{{\cal I}}
\int \frac{d^{4}x}{(2\pi)^{4}} \frac{d\hat{T}d\hat{T}^{'}}{4\pi}
Y_{II_z}^{*}(\hat{T})Y_{II_z}(\hat{T}^{'})\hat{T}_{\mu}\hat{T}^{'}_{\mu}
e^{iK\cdot x} \rho_{\pi}(x)
(e^{ \rho_{\pi}(x)\hat{T}\cdot\hat{T}^{'}+\rho_{\omega}(x) }-1),
\label{eq:pionNumber}
\end{equation}
and the mean number of $\omega$'s by
\begin{equation}
N_{\omega} =\frac{1}{{\cal I}}
\int \frac{d^{4}x}{(2\pi)^{4}} \frac{d\hat{T}d\hat{T}^{'}}{4\pi}
Y_{II_z}^{*}(\hat{T})Y_{II_z}(\hat{T}^{'})
e^{iK\cdot x} \rho_{\omega}(x)
(e^{ \rho_{\pi}(x)\hat{T}\cdot\hat{T}^{'} +\rho_{\omega}(x)
}-1).
\label{eq:omegaNumber}
\end{equation}
These can again be calculated using the expansion method.
It is clear that correlations can be calculated as before \cite{Amado2},
but we will not do so here.

\section{Results}
Our  calculational plan is now clear.  We begin with a
spherically symmetric ``blob" of Skyrmionic matter of
size fixed by the $\pi$-meson mass and amplitude fixed
by a total energy of 2 GeV, but with baryon number zero.  We
use Skyrme dynamical equations modified to include the omega field
to propagate the classical pion and omega fields outward into
the radiation zone. These classical radiation fields are then
used to construct quantum coherent states for the $\pi$-mesons
and for the $\omega$-mesons, and the coherent states are
projected onto states of good isospin and four-momentum.
We use these states to find mean meson numbers. Note that
except for the paramters in the Skyrme model, and these are
fixed by nucleon physics, there are no free parameters in
our picture.

We first calculate mean meson numbers from the purely classical
fields. This is completely equivalent to calculating the mean
numbers in the coherent state without the isospin and four-momentum
projections. We find
$$
N_{\pi}=4.4
$$
$$
N_{\omega}=0.86
$$
$$
N_{\mbox{total}}=7.0
$$
where $N_{\mbox{total}}$ is the total number of pions, coming
from both direct pions and from omega decay. It is given by
$N_{\mbox{total}} = N_{\pi} + 3 N_{\omega}$.  Our case
corresponds to a percentage of secondary pions of $ 37\%$.
Note that the mean number of omega mesons is quite small.
It is usual to say that coherent state methods are best
applied in the limit of a large number of quanta, and
$0.86$ is certainly not a large number.  Nevertheless we use
coherent states methods for this case both to remain parallel
with the treatment of the pions, and because we have little
idea of what else to do. An after-the-fact justification comes
from the close agreement with data.

We now turn to a calculation of the number of pions and
omega meson from annihilation
with both isospin and
four-momentum projections included as in (\ref{eq:pionNumber}) and
(\ref{eq:omegaNumber}).
In the table we show the number of pions (now separated by
charge type) and number of omegas for each of the isospin
channels that can be reached in nucleon-antinucleon annihilation
at rest.
We also show the total number of pions of each charge
type as well at the total number of all types.  These
are calculated combining the direct pions with those from
omega decay using the decay mode $\omega \rightarrow
\pi^+ + \pi^0 + \pi^- $.  We also give the
pion variance for each channel.
We see, as we expect from its large
mass, that the mean number of omega mesons is decreased by
the imposition of energy and momentum conservation.
The mean number of pions, the
variance, and the fraction by charge
type is quite close to what is observed \cite{Amsler}.
The percentage of
secondary pions is $23\%$ for all three isospin channels.  This
too is roughly what is seen when allowance is made for pions
from rho mesons, that we have not yet included.  It may by
argued that since the mean number of pions comes out about
right, there is no room for pions from rho mesons.  That
argument is incorrect, because it
neglects the important effect of energy conservation. Note
that including omega mesons has not substantially raised the mean
number of pions from what we obtain in our earlier, pions only,
calculation \cite{Amado1},\cite{Amado2}.
Because the total energy is fixed, when we add degrees of
freedom, the energy is redistributed, less goes into direct pions
and more into indirect, but the total remains about the same.
We expect this will be true when rho mesons are included, but that
then the number of secondary pions will be close to the
experimental value of $40\%$.

The pion averages are interesting, but the pion number spectrum is a
more stringent test of the formalism. We are interested in the
probability, $ P_n$, of finding exactly $n$ pions
in annihilation at rest, from states of fixed isospin and four-momentum.
This can be calculated from the
joint  probability of finding $N_{\pi}$ pions, and $N_{\omega}$
omega mesons, $p(N_{\pi},N_{\omega})$
using the relation
\begin{equation}
 P_n = \sum_{n=N_{\pi}+3N_{\omega}} p(N_{\pi},N_{\omega}). \label{eq:piondist}
 \end{equation}
 The joint probability, $p(N_{\pi},N_{\omega})$,
 is given by Eq.(\ref{eq:jointProb}).
 This is the generalization to two meson
 types of the method used for pions only in \cite{Amado2}.
 In Figures 11, 12 and 13, we show $P_n$ as a function of $n$ for each of the
 three isospin cases.  The solid squares are the results of our
 calculation, the open circles are a normalized gaussian distribution
 fixed to the same average and variance as our calculation.  We see,
 as we saw in our previous work \cite{Amado1}, \cite{Amado2}, that
 with four-momentum conservation imposed, our calculation is
 indistinguishable from a gaussian distribution, even though our
 reaction mechanism is very far from a statistical one.  The pion
 number distributions in the three figures look very much like
 the empirical ones \cite{Dover}. In our previous, pions only, calculation
 \cite{Amado2}, we found that $I=0$ states could go only into
 even numbers of pions and $I=1$ states into odd numbers.
 No such effect is seen in the data. Note that by
 including the omega meson we have smoothly removed any
 vestige of this even-odd effect.  All of this gives futher
 support to the quantized classical wave picture of annihilation.

\section{Conclusion}
We have shown that a classical treatment of nucleon-antinucleon
annihilation at rest based on Skyrme dynamics extended to
include the omega meson and then quantized using coherent
states gives an excellant account of the pion spectra
seen in annihilation.  This treatment also removes an
odd-even artifact of the pions only treatment we presented
before \cite{Amado1}, \cite{Amado2}. The treatment
presented here also is more firmly based in the classical
dynamics and hence more rooted in QCD. In fact our
treatment now has no free parameters, and yet fits the
major trends of the data very well.  Its major failing is
that we find only 23\% of annihilations going via vector mesons,
while empirically that number is closer to 40\%.
This missing fraction is presumably due to annihilations into
rho mesons, which are not yet in our picture. We argue that
including them will boost the fraction of annihilations into
vector mesons while enery-momentum conservation will prevent
that boost from spoiling argeement with pion number distributions.

Our approach to annihilation is part of a  widening family of
approaches to strong interaction
physics problems that fall squarely in the domain of
non-perturbative QCD, but that involve fairly large energy release.
These problems are very difficult to solve in full QCD.  The alternative
is to solve them first in classical QCD (CCD). There the dynamics
is difficult but often tractable.  In our case it is the Skyrme
model modified to include omega mesons.  Once the classical fields
have been obtained, they can be quantized in the radiation zone
using the method of coherent states and some of the important
quantum numbers can even be imposed.  These quantized states
can then be compared with experiment.
In our  case we find good agreement, as far as we have gone, and are
encouraged to go further.  In particular we will study pion
correlations, annihilation in flight, the effect of the rho meson
and related phenomena.  It should  be noted that one of the benefits
of the classical dynamics-coherent state approach is that all final
channels are treated together. Further afield this direction,
using classical QCD first and quantizing later,  is being
applied to disoriented chiral condensates,
Centauro events \cite{Bjorken},
\cite{Kowalski}, \cite{Wilczek}, \cite{Blaizot}, \cite{Gavin},
\cite{Kogan}, and
much else.  It may have other uses in high energy heavy ions
collisions and in other problems of coherent hadronization.

\section*{Acknowledgements}
This work is supported in part by the United States
National Science Foundation.

\newpage
\begin{flushleft}
Figure Captions \\
\vspace{1in}
Figure 1. Pion field configuration F as a function of r and t. \\
\vspace{0.3in}
Figure 2. $\omega_{0}$ (in units of omega mass m)
field as a function of r and t. \\
\vspace{0.3in}
Figure 3. $\omega_{r}$ (in units of omega mass m)
field as a function of r and t. \\
\vspace{0.3in}
Figure 4. Total energy density multiplied by $4\pi r^{2}$
(in units of MeV/fm) as a function of r and t. \\
\vspace{0.3in}
Figure 5. Pion energy density multiplied by $4\pi r^{2}$
(in units of MeV/fm) as a function of r and t.  \\
\vspace{0.3in}
Figure 6. Energy density associated
with $\omega_{r}$ multiplied by $4\pi r^{2}$
(in units of MeV/fm) as a function of r and t. \\
\vspace{0.3in}
Figure 7. Interaction energy density multiplied by $4\pi r^{2}$
(in units of MeV/fm) as a function of r and t. \\
\vspace{0.3in}
Figure 8. Pion momentum distribution. The horizontal axis
is the scaled momentum
in units of inverse omega mass. The vertical axis is the dimensionless
reduced pion
distribution function. The area under the curve gives the total
direct pion number $N_{\pi}$. \\
\vspace{0.3in}
Figure 9.
Omega momentum distribution. The horizontal axis is the scaled momentum
in units of inverse omega mass. The vertical axis is the dimensionless
reduced omega
distribution function. The area under the curve gives the total
omega number $N_{\omega}$. \\
\vspace{0.3in}
Figure 10. Pion momentum distribution.
Solid line: Skyrme calculation; Dot-Dashed Line: Phenomenological
pion momentum distribution used in [1,2].
The horizontal axis is the scaled momentum
in units of inverse omega mass. The vertical axis is the dimensionless
reduced  pion
distribution function. The area under the curve gives the total
pion number $n$.  \\
\vspace{0.3in}
Figure 11. Pion number distribution in $I=0, I_z=0$ channel.
Solid squares are given by the distribution $P_{n}$.
Circles are the gaussian distribution with the same mean and variance.  \\
\vspace{0.3in}
Figure 12. Pion number distribution in $I=1, I_z=0$ channel.
Solid squares are given by the distribution $P_{n}$.
Circles are the gaussian distribution with the same mean and variance. \\
\vspace{0.3in}
Figure 13. Pion number distribution in $I=1, I_z=1$ channel.
Solid squares are given by the distribution $P_{n}$.
Circles are the gaussian distribution with the same mean and variance. \\
\end{flushleft}

\begin{table}
\caption{pion number distribution. $N_{\pi}$ is the total direct
pion number and $N_{\omega}$ is the total omega number.
$n_{0}$ is the total $\pi^{0}$ number including those from omega decay.
Similarly $n_{+}$ and $n_{-}$ are the total $\pi^{+}$ and $\pi^{-}$
numbers including those from omega decay. $n$ is the total pion number
and $\sigma$ is the standard deviation for $n$ calculated from
the distribution function $P_{n}$.}
\begin{center}
\begin{tabular}{|c||ccccccc|}   \hline
Channel & $N_{\pi}$ & $N_{\omega}$ &
$n_{0}$  & $n_{+}$  & $n_{-}$  & $n$ &  $\sigma$   \\  \hline\hline
$I=0\hspace*{2.0ex}I_z=0$ & 4.94 & 0.60
&  2.25  &  2.25  &  2.25  &  6.75  & 0.79  \\ \hline
$I=1\hspace*{2.0ex}I_z=0$ &  5.17 & 0.48 &
3.99  &  1.31  &  1.31  &  6.61  & 0.84  \\ \hline
$I=1\hspace*{2.0ex}I_z=1$ & 5.20 & 0.48
&  1.32  &  3.16  &  2.16  &  6.64  & 0.84
\\ \hline
\end{tabular}
\end{center}
\end{table}

\end{document}